\documentstyle[12pt]{article}
\input epsf

\begin{document}

\begin{titlepage}

\begin{flushright}
Freiburg--THEP 96/20
October 1996
\end{flushright}
\vspace{2.5cm}

\begin{center}
\large\bf
{\LARGE\bf On the position of a heavy Higgs pole}\\[2cm]
\rm
{Adrian Ghinculov and Thomas Binoth}\\[.5cm]

{\em Albert--Ludwigs--Universit\"{a}t Freiburg,
           Fakult\"{a}t f\"{u}r Physik}\\
      {\em Hermann--Herder Str.3, D-79104 Freiburg, Germany}\\[3.5cm]
      
\end{center}
\normalsize

\begin{abstract}
Higher loop calculations in the Higgs sector of the standard model 
at the Higgs mass scale have shown that perturbation theory diverges very
badly at about 1 TeV in the on--shell renormalization scheme. 
The prediction of the position of the Higgs pole in the complex $s$ plane 
becomes unreliable. We show that in the pole
renormalization scheme this appears to have much better convergence properties,
while showing good agreement with the 
on--shell scheme over the validity range of the latter. 
This suggests that the pole scheme should be preferable 
for phenomenological studies of heavy Higgs bosons.

We discuss whether
this behaviour can be the result of a certain relation between the on--shell
mass and the pole mass at the nonperturbative level.
\end{abstract}


\end{titlepage}


\title{On the position of a heavy Higgs pole}

\author{Adrian Ghinculov and Thomas Binoth}

\date{{\em Albert--Ludwigs--Universit\"{a}t Freiburg,
           Fakult\"{a}t f\"{u}r Physik},\\
      {\em Hermann--Herder Str.3, D-79104 Freiburg, Germany}}

\maketitle

\begin{abstract}
Higher loop calculations in the Higgs sector of the standard model 
at the Higgs mass scale have shown that perturbation theory diverges very
badly at about 1 TeV in the on--shell renormalization scheme. 
The prediction of the position of the Higgs pole in the complex $s$ plane 
becomes unreliable. We show that in the pole
renormalization scheme this appears to have much better convergence properties,
while showing good agreement with the 
on--shell scheme over the validity range of the latter. 
This suggests that the pole scheme should be preferable 
for phenomenological studies of heavy Higgs bosons.

We discuss whether
this behaviour can be the result of a certain relation between the on--shell
mass and the pole mass at the nonperturbative level.
\end{abstract}


\section{Introduction}

The divergent nature of the perturbation theory can be particularly
disturbing especially when the coupling constant is large
and the divergent behaviour sets in at low order in the loop expansion.
Apart from the fundamental problem of disentangling the physical
information out of a divergent perturbative expansion, 
on the phenomenological side one may find out that one is unable
to make quantitative predictions of sufficient accuracy because 
the first few radiative corrections are large, after which 
the divergent behaviour sets in.

In the hope of elucidating the problem of the electroweak symmetry 
breaking mechanism, the LHC will be able to search for a Higgs 
boson up to masses of the order of 1 TeV. A number of higher order 
calculations of processes at the Higgs resonance became available recently. 
They indicate that for such heavy Higgs bosons the higher order radiative 
corrections become indeed large \cite{q96}. 
In the on--shell renormalization scheme, 
if the Higgs mass is larger than 930 GeV, the two--loop correction 
to the Higgs decay into vector bosons exceeds the one--loop correction 
\cite{2loop:Htoww, frink}. 
For the Higgs decay into fermions, this happens at about 1.1 TeV
\cite{2loop:Htott,kniehl}. 
Similar conclusions are valid for the 
$gg \rightarrow H \rightarrow ZZ$ process
\cite{glufusion},
which will be tested at the LHC, and for the 
$f\bar{f} \rightarrow H \rightarrow  f^{\prime} \bar{f^\prime}$ 
and
$f\bar{f} \rightarrow H \rightarrow  ZZ$ 
scatterings
\cite{glufusion,2loop:method}, 
which were proposed as a production mechanism for Higgs bosons 
at a possible muon collider.

If the one-- and two--loop corrections which are available are not 
accidentally very small or very large, and are indicative for the 
divergent behaviour of the perturbation series, then one must conclude
that beyond these limits -- of the order of 900---1000 GeV, 
depending on the process -- the perturbation theory is totally 
unreliable in the on--shell scheme. Even for lower masses, the theoretical 
uncertainty due to the unknown higher order corrections may be 
substantial. This view is supported by calculations involving 
resummations of higher order logarithmic contributions 
\cite{riesselmann}, where 
by examining the scheme dependence of the results one can estimate 
the size of the unknown higher order corrections.

How to recover a physical prediction out of a divergent and not 
Borel summable perturbative expansion is still an open question. 
Considerable progress has been made in understanding the 
large order behaviour of perturbation theory, and this may lead to 
recipes for summing up the perturbative expansion. As an example, 
the commensurate scale relations among effective charges show promise
of dealing with the factorial behaviour associated with the 
infrared renormalon structure of QCD 
\cite{brodsky}. The justification is
that this factorial growth is anyway related to the unknown 
behaviour of the beta function in a region where the perturbative
solution in unreliable.

As it happens, {\em worsening} the divergent behaviour of a perturbative 
expansion is a much easier task. One way of doing this is suggested
by ref. \cite{beenakker}. Suppose one starts with a perturbative
expansion in a given renormalization scheme, with an expansion parameter 
$\lambda$. Suppose that this series converges well, up to a high enough 
loop order, for an expansion parameter $\lambda$ smaller than a critical 
value $\lambda_c$. In this scheme the physics corresponding to values of 
$\lambda$ smaller than $\lambda_c$ are described well, with controllable 
accuracy. In order to worsen the divergency of the perturbation 
series, one can define a new renormalization scheme by choosing a new 
expansion parameter $\bar{\lambda}$, so that the 
exact, nonperturbative relation between the two expansion parameters, 
$\bar{\lambda} = \bar{\lambda}(\lambda)$,
has a cut starting at $\lambda = \lambda_c^\prime$.
$\lambda_c^\prime$ can be chosen to be conveniently smaller 
than $\lambda_c$. Of course,
then the power expansion in terms of the new expansion parameter 
$\bar{\lambda}$ will converge satisfactorily only over a smaller 
onset of physics than the perturbation theory in the original scheme.
For instance, one can arrange things so that
the renormalization conditions in the new scheme 
have a solution only for a limited onset 
of physics, corresponding to  $\lambda < \lambda_c^\prime$. 

Of course, this also may happen accidentally, if one performs a calculation
in an inconvenient renormalization scheme. In ref. 
\cite{beenakker} it was argued that under certain assumptions
the on--shell renormalization conditions for a boson propagator 
may have no solution for a range of physics for which 
the pole mass and width are well defined. To show this, a model was 
considered which allows an all--order, nonperturbative solution. 

If the exact solution is unknown, and one only knows a few orders in 
perturbation theory, as is the case with the standard model Higgs sector, 
it is more difficult to establish unambiguously 
whether such a mechanism is indeed present. 
For instance, perturbation theory in the on--shell scheme only gives the 
Higgs propagator as an expansion in the on--shell Higgs mass $m_H$, and
one cannot directly decide whether the divergent behaviour which one 
observes is mainly due to a strong selfinteraction of the Higgs field
or if there is a limit of the values of the on--shell mass $m_H$ 
beyond which the on--shell renormalization conditions have no solution.
If that was the case, larger values of $m_H$ would be unphysical, however
stronger selfcouplings than this limit may be possible, and may 
be appropriately described in the pole renormalization scheme.

In the following section we review the existing knowledge on the 
position of a heavy Higgs pole coming from perturbation theory 
in the on--shell scheme. We then derive the corresponding result 
in the pole renormalization scheme, and show that it converges much better.
We analyze the nonperturbative $1/N$ expansion of an $O(N)$ model.
We find no branching point in the $m_H=m_H(M)$ relation
at leading order which could explain the observed divergency pattern, 
and speculate on the possibility that such a branching 
point may be induced at ${\cal O}(1/N)$.


\section{The Higgs pole in the on--shell scheme}

We are interested in effects related to a heavy Higgs boson, so
we adopt the framework proposed in ref. 
\cite{marciano} 
for calculating leading
effects in the Higgs mass. This reduces to considering the 
Higgs--Goldstone Lagrangian of the standard model in Landau gauge.

The Higgs propagator receives quantum corrections which 
lead to a momentum dependent self--energy. In the on--shell scheme,
the Higgs propagator reads:

\begin{equation}\label{prop}
  P(s) = \frac{i}{s - m_H^2 + \Sigma(s)}
     \; \; \; \; \; \; ,
\end{equation}
and the on--shell mass is defined by the renormalization condition:

\begin{equation}
  {\cal R}e \left( i P^{-1}(m_H^2) \right) = 0
     \; \; \; \; \; \; .
\end{equation}

The quartic coupling $\lambda$ is related at all orders 
to the on--shell mass $m_H$ by the relation  
$\lambda=G_F/(2\sqrt{2}\pi^2)\, m_H^2 \equiv c\, m_H^2$,
were $G_F = 1.16637 \times 10^{-5}$ GeV$^{-2}$ is the Fermi constant.

The self--energy $\Sigma(s)$
is given in perturbation theory by a power series 
with the expansion parameter $\lambda$. 
Its real part has no constant or linear terms in $s$,
since these are absorbed by the mass and wave function renormalization.

The propagator in eq. 1 has a complex pole at an energy $s_P$ which is
given by the equation $ P^{-1}(s_P) = 0$. The pole mass and width are 
then defined by $s_{P} = ( M -i\Gamma/2 )^2 $.
Clearly, the on--shell mass $m_H$ and the pole mass $M$ are not the same.

In the on--shell scheme, one can express $M$ and $\Gamma$
as power series in the expansion parameter $m_H$.
For calculating the location of the Higgs pole one has to solve the equation
$ P^{-1}(s_P) = 0$ in the second Riemann sheet of the complex $s$ plane. 
It is easy to do this at the one--loop order, where the analytical
expression of the self--energy is trivial. We are interested in including the
two-- and three--loop contributions as well, and for some of these 
contributions only on--shell results have been calculated so far. 
Nevertheless, one can solve
the pole equation in the complex plane up to the desired order if enough 
on--shell information is available.

To do this, it is useful to double expand
the self--energy in the coupling constant $\lambda$
and in the energy distance from the on--shell mass $s-m_H^2$:

\begin{eqnarray}\label{expand}
\Sigma(s) &=& A + (s-m^2_H) \,B + (s-m_H^2)^2 \,C + \dots \\ &&\nonumber \\
A &=& i \,{\cal I}m\Sigma(m_H^2)\, = 
      i\,m_H^2 \Bigl(a_1\lambda+a_2\lambda^2+a_3\lambda^3+a_4\lambda^4+\dots\Bigr)\nonumber\\
B &=& i \,{\cal I}m\Sigma'(m_H^2) =  
      i\,
       \Bigl(b_1\lambda+b_2\lambda^2+b_3\lambda^3+\dots\Bigr)\nonumber\\
C &=& \Sigma''(m_H^2)/2 \quad = \, m_H^{-2} \Bigl( c_1\lambda +c_2\lambda^2 +\dots \Bigr) \nonumber
\end{eqnarray}

Here, $a_j$, $b_j$ are real valued, whereas $c_j$ are complex in general.
The corrections to the Higgs decay width are known at two--loop order,
so $A$ is known with three--loop precision.
$B$ is known to two--loop, and $C$ is known to vanish at one--loop.

The terms in the expansion which we need for finding a consistent solution
of the pole equation at three--loop level are 
\cite{2loop:Htoww}---\cite{jikia}:

\begin{eqnarray}
a_1 = 3\pi/8 &,& a_2 = a_1\cdot 0.350119 \quad ,\, a_3 = a_1 \cdot (0.97103 + 0.000476) \nonumber\\
b_1 = 0  \hspace{0.7cm} &,& b_2 = 1.002245 \qquad \,\,\,\,\,,\, c_1 = 0.2181005 
\end{eqnarray}

Indeed, one can convince oneself by solving directly the pole equation
up to order $\lambda^4$ that its solution reads:

\begin{eqnarray}
s_P = m_H^2 \left[ 1 - i (a_1 \lambda + a_2 \lambda^2 + a_3 \lambda^3) 
                     - ( a_1 b_2 - a_1^2 c_1 ) \lambda^3 \right] 
+  {\cal O}(\lambda^4)
\end{eqnarray}
and therefore the higher order unknown coefficients $a_4$, $b_3$ and $c_2$
are not needed for solving the pole equation consistently with three--loop 
accuracy.

One further obtains:

\begin{eqnarray}
  \sqrt{s_P} & = & m_H -i\, \frac{c}{2} a_1 m_H^3
                          + \frac{c^2}{8} (a_1^2 - 4 i a_2 ) m_H^5 
        \nonumber\\
             &   & + \frac{c^3}{16} \left[(4 a_1a_2 - 8 a_1 b_2 + 8 a_1^2 c_1) 
	           + i (a_1^3-8a_3)\right] m_H^7 
     \; \; \; \; \; \; ,
\end{eqnarray}
so that the pole mass and width read:
 
\begin{eqnarray}
  M    &=& m_H + \frac{c^2a^2_1}{8}m_H^5 
               + \frac{c^3(a_1a_2-2a_1b_2 + 2 a_1^2 c_1)}{4} m_H^7 
        \nonumber\\
\Gamma & = & c\, a_1 m_H^3 + c^2 a_2 m_H^5 
                           + \frac{c^3 (8a_3-a_1^3)}{8} m_H^7 
     \; \; \; \; \; \; .
\end{eqnarray} 

\begin{figure}[t]
\hspace{1.5cm}
    \epsfxsize = 15cm
    \epsffile{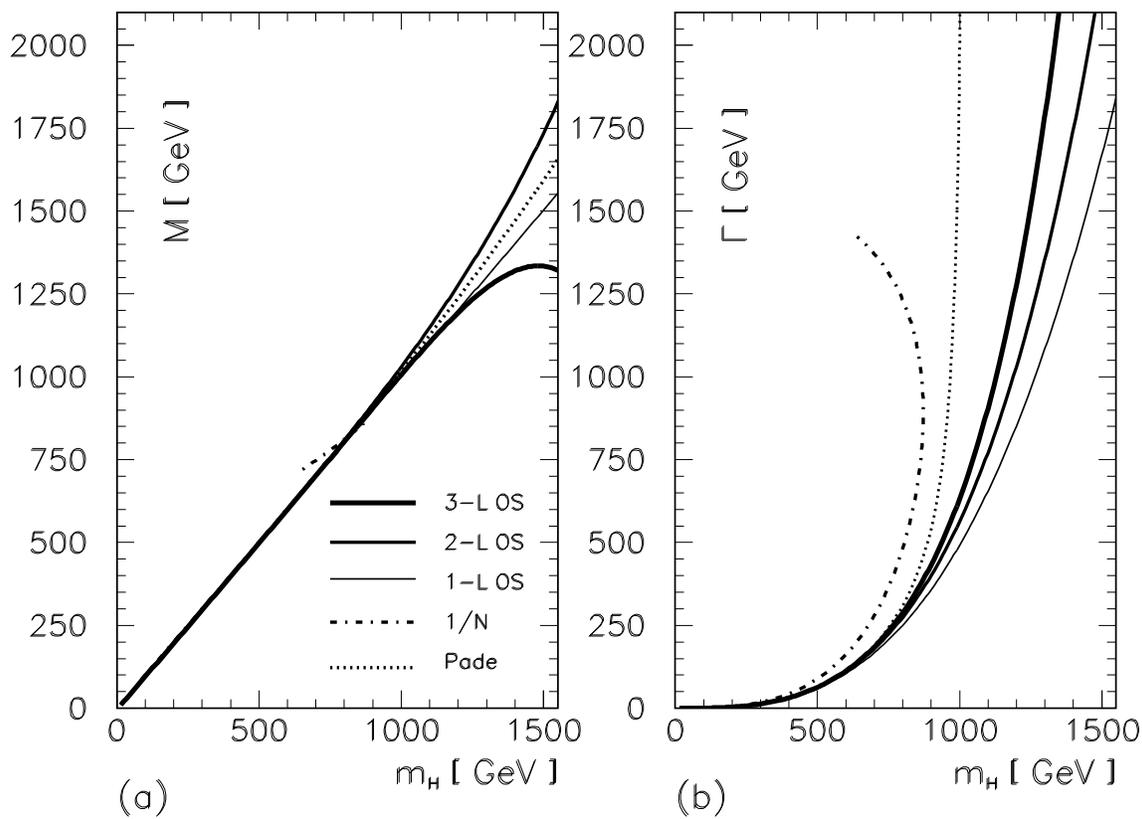}
\caption{{\em The real and the imaginary parts of the position of the Higgs
         pole at LO, NLO and NNLO in the on--shell perturbative expansion.
         In fig. a), the $1/N$ result practically coincides with all
         the other curves up to about $m_H=872$ GeV. Its behaviour is
         shown in detail in fig. 3.
         For comparison, the 
         $[2/1]$,  $[1/2]$ (a) and the  $[1/1]$ (b) Pad\'e approximants of 
         the perturbation series are also shown, 
         as well as the LO $1/N$ result. The Pad\'e approximants 
         $[2/1]$,  $[1/2]$ of fig. (a) practically coincide 
         for this range of $m_H$.}}
\end{figure}

The above relation between the pole and the on--shell masses agrees
with the result derived in ref. \cite{willenbrock}.

We plot eqns. 7 in fig. 1. For comparison, we show also the
$[2/1]$,  $[1/2]$ (a) and the  $[1/1]$ (b) 
Pad\'e approximants of eqns. 7, as well as the leading order of the
nonperturbative $1/N$ expansion, which will be discussed in the following.
The two--loop correction to the pole width equals the one--loop correction 
at about 1 TeV, similar to the results of. ref. \cite{2loop:Htoww,frink}. 
The corrections 
to the pole mass appear to be somehow smaller. The $[2/1]$ and $[1/2]$ 
Pad\'e approximants
of fig. 1 a) do not improve much the agreement of perturbation theory
with the $1/N$ expansion for large couplings; 
the $[1/1]$ approximant of fig. 1 b) shows 
qualitatively a saturation of the mass similar to the $1/N$ result,
but the numerical discrepancy remains considerable even for masses 
as low as 600---650 GeV.

The position of the pole in the on--shell scheme, which is given by
eq. 6, is shown in fig. 2, along with the $1/N$ expansion and
the pole renormalization scheme result, which will be derived in the following.
We have marked in fig. 2 the point beyond which  
the NNLO correction to eq. 6 exceeds the NLO correction in absolute value. 
This corresponds to an on--shell mass of about 980 GeV, but the perturbative
series may be untrustworthy long before.
Of course, other measures of the degree of divergency of the series are
possible -- for instance the point beyond which 
the distance between the NNLO and 
the NLO results is larger than that between the NLO and the LO results,
for a given pole mass. This happens for a mass larger than about 710 GeV.


\section{The Higgs pole in the pole scheme}

In the pole renormalization scheme, the Higgs propagator reads:

\begin{equation}
P(s) = \frac{i}{s - M^2 + \Sigma(s)}
\end{equation}

The coupling of the theory is parameterized by the pole mass $M$, 
which is defined by the condition:

\begin{equation}
  P^{-1}((M - \frac{1}{2} \Gamma)^2)  = 0
\end{equation}

The Higgs width is then expressed as an expansion in $M$.

We cannot solve directly the renormalization conditions 
at three--loop order because
the analytical continuation of the self--energy in the second Riemann sheet
is not available. Nevertheless, the relation between the observables 
$M$ and $\Gamma$ is independent of some intermediary renormalization 
scheme in which they are calculated. In the on--shell scheme, $M$ and 
$\Gamma$ are given by eqns. 7, so one can invert the $m_H$ power series
of $M$ to obtain $m_H$ as a power expansion in $M$, and substitute 
in the second line of eqns. 7, 
for obtaining the following pole renormalization scheme result:

\begin{equation}
  \Gamma = M \left[ a_1 c M^2 + c^2 a_2 M^4 
                              + c^3 \Bigl( a_3 - a_1^3/2\Bigr) M^6 \right]   
\end{equation}

This relation is plotted in fig. 2, along with the on--shell result and
the prediction of the leading order $1/N$ expansion. The interesting point
is that the pole scheme expansion appears to converge much better than 
the on--shell scheme. In eq. 10, the NLO and the NNLO corrections 
become equal for $M = 1.74$ TeV. At the same time, the NNLO predictions
in the on--shell and the pole schemes agree very well over the energy
range where the former is supposed to be a good approximation. However,
the $1/N$ expansion deviates considerably from the perturbative solution
already at 600---650 GeV, where perturbation theory should provide a
reliable result.

\begin{figure}
\hspace{1.5cm}
    \epsfxsize = 14cm
    \epsffile{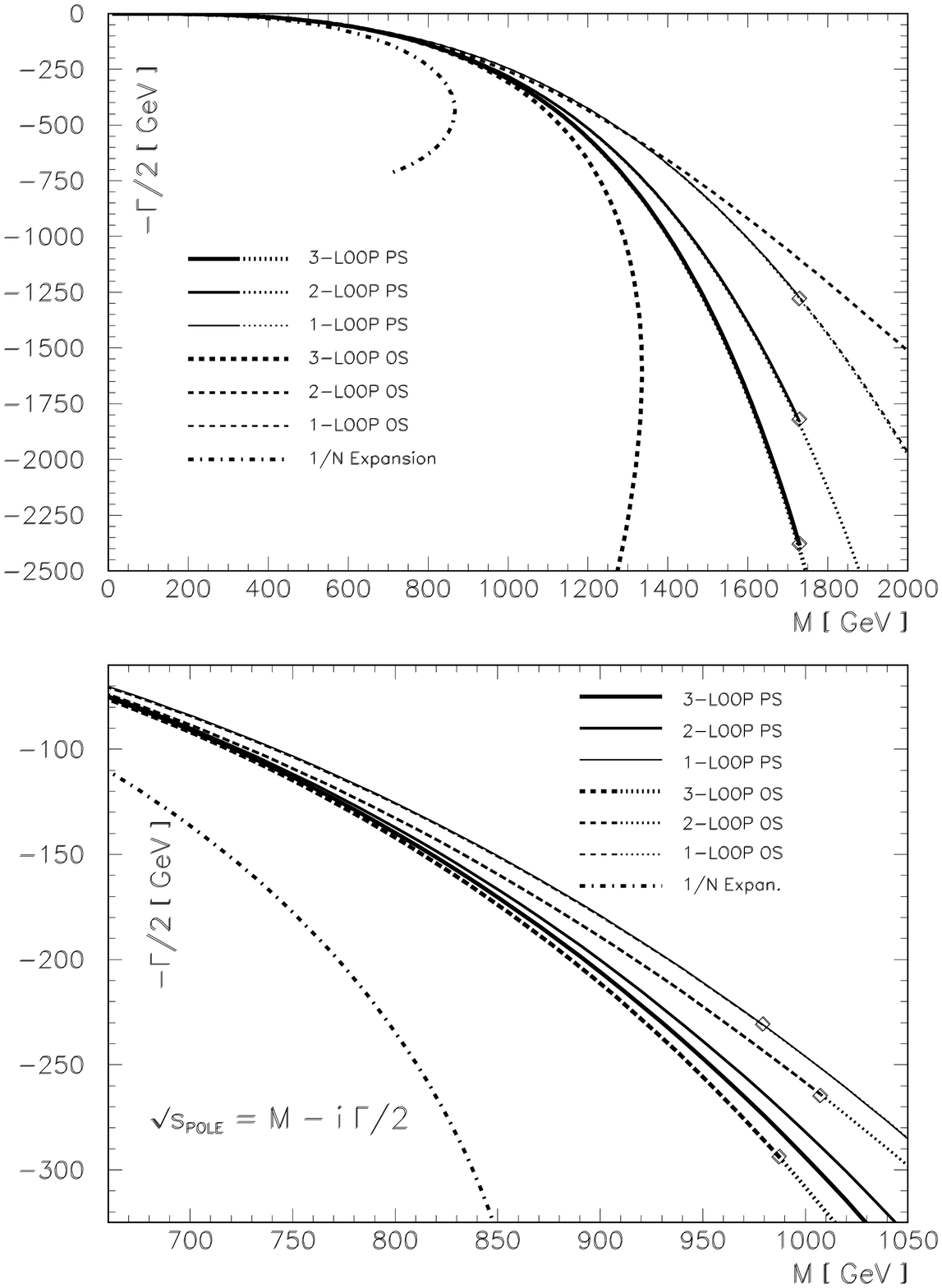}
\caption{{\em The position of the Higgs pole in the complex 
         $\protect\sqrt{s}$ plane. We show the on--shell scheme (OS) result 
         of eq. 6, the pole scheme (PS) result of eq. 10, and the leading order
         $1/N$ expansion result of eqns. 14. The points beyond which the
         NNLO correction is larger than the NLO correction are marked
         for the pole scheme (top) and for the on--shell scheme (bottom).}}
\end{figure}

If the size of the NLO and NNLO corrections in the pole scheme  
is representative for the 
divergency of the perturbative series, 
this behaviour suggests that the pole scheme is a better framework for 
describing a heavy Higgs for phenomenological purposes. 


\section{Beyond perturbation theory}

A number of recipes exist in QCD for finding an appropriate renormalization
scheme, in which the perturbative results ought to convergence better. 
The accuracy of the calculation can be spoiled by choosing a very different
renormalization scale because of the presence of large logarithms in higher
orders.  

The perturbative expansion at the Higgs energy scale appears to
converge much better in the pole renormalization scheme than in
the on--shell scheme, but no large logarithms can be made responsible 
for this. In this section we address the question whether 
this feature may be due to the existence at the nonperturbative level 
of a certain relation between the on--shell and the pole masses. 

The convergence
pattern which is observed in the on--shell versus the pole scheme
can be understood if one assumes that there is a  nonperturbative relation
between $m_H$ and $M$, $m_H = m_H(M)$, which has a branching point 
at a value $M_c$ of the order of 1 TeV. Of course, this would induce
a strong divergent behaviour in the on--shell scheme near the singularity,
even if there the Higgs field would not be truly strong selfinteracting. 
Beyond that value, the on--shell mass ceases to be a good parameterization
of the Higgs selfcoupling. Nevertheless, the quartic coupling of the Higgs 
field may still be not very strong, and the physics at energies comparable
to the Higgs mass may be described appropriately in the pole scheme.

The authors of ref. \cite{beenakker} considered a model of a W boson coupled 
to a large number of light fermions, which is an exactly solvable model,
and which displays such a relation between the on--shell and the 
pole masses. It is more difficult to establish whether a similar 
scenario is indeed present in the standard model. To gain further 
insight, one has to go beyond the standard perturbation theory. 
One promising approach is the nonperturbative $1/N$ expansion of 
an $O(N)$ sigma model. We anticipate that the leading order solution 
of the $1/N$ expansion does not display the type of relation between 
$m_H$ and $M$ we are looking for. Nevertheless, such a mechanism  
may be generated at next--to--leading order.

For fixing the notations, we consider an $O(N)$ sigma model with the
Lagrangian:

\begin{equation}
  {\cal L} =   \frac{1}{2} (\partial_{\mu} \Phi)(\partial^{\mu} \Phi) 
             + \frac{1}{2} \mu^2 \Phi^2
	     - \frac{\lambda}{N} \Phi^4
	     \; \; \; \; \; .
\end{equation}

One performs the calculation as an expansion in $1/N$, keeps only 
the leading order, and sets in the end $N=4$ and the vacuum expectation 
value of the Higgs field $v$ to 246 GeV. We merely quote the result for 
the Higgs propagator, without repeating this well--known summation 
of Goldstone loops \cite{einhorn}:

\begin{equation}\label{1vsn}
  P(s) =  i  \left[ s - \frac{2\lambda v^2}{1-\frac{\lambda}{4\pi^2}
                       \Bigl(\log(s/\mu^2)-i\pi \Bigr)} \right]^{-1}
\end{equation}

In this expression, the divergency of the bubble diagrams was absorbed 
in the renormalization of the coupling constant $\lambda$, and one still 
has the freedom to perform a finite renormalization. Of course, 
the relation between the Higgs mass and width is independent of 
the actual renormalization scheme.

Following Einhorn \cite{einhorn}, we will define the coupling constant at the
energy scale of the Higgs pole, $\mu^2 = |s_P|$, which leads to a convenient parameterization 
of the results. Therefore, with the definitions:

\begin{eqnarray}
  s_P  & = &  \left( M - \frac{i}{2} \Gamma \right)^2  
              \; = \; \mu^2 e^{- 2 i \theta}
      \nonumber\\
   x   &= & \tan(\theta)   \; \;\; \;\; \; , \; \;	x \in (0,1)     
  \; \; \; \; \;\; \; \; \; \; ,
\end{eqnarray}
one obtains the following parameterization:

\begin{eqnarray}
  M      & = &  \frac{4 \pi v}{\sqrt{\pi + 2 \arctan(x)}} 
                \frac{\sqrt{x}}{1 + x^2}
      \nonumber\\
\Gamma   &= &   \frac{8 \pi v}{\sqrt{\pi + 2 \arctan(x)}} 
                \frac{x \sqrt{x}}{1 + x^2}     
      \nonumber\\
\lambda  &= &   \frac{8 \pi^2}{\pi + 2 \arctan(x)} 
                \frac{x}{1 - x^2}     
      \nonumber\\
\mu^2    &= &   \frac{16 \pi^2 v^2}{\pi + 2 \arctan(x)} 
                \frac{x}{1 + x^2}     
  \; \; \; \; \;\; \; \; \; \; .
\end{eqnarray}

For each value of $x$, which measures the coupling strength of the
Higgs field, the on--shell Higgs mass $m_H$ is given by the following
transcendental equation:

\begin{equation}
  m_H^2 \left\{ \left[ 1 - \frac{\lambda}{4 \pi^2}
                          \log(\frac{m_H^2}{\mu^2}) \right]^2 + 
               \left( \frac{\lambda}{4 \pi} \right)^2 \right\}
   =
  2 v^2 \lambda 
  \left[ 1 - \frac{\lambda}{4 \pi^2}\log(\frac{m_H^2}{\mu^2})  \right]
  \; \; \; \; \; \; \; ,
\end{equation}
with $\lambda$ and $\mu^2$ given by eqns. 14.

At low values of the coupling, the pole mass and the on--shell mass
have practically the same value, but start to deviate for larger couplings.
As the coupling $x$ increases, the pole mass saturates and then starts
to decrease, while the width continues to grow. The pole mass $M$ reaches 
its maximum of 867 GeV for $x = 0.501$. The on--shell mass has a similar 
behaviour.  Its maximum of 872 GeV is reached for a slightly larger 
value of the coupling, $x = 0.515$. 

The relation between the pole and the on--shell masses is shown 
in fig. 3. It shows that in this approximation both masses are 
practically equally good parameterizations of physics. They reach their 
maxima at nearly the same values of the coupling, and for any value 
of the coupling $x \in (0,1)$ both on--shell and pole renormalization 
conditions have solutions. No branching point is present in the exact 
relation between $m_H$ and $M$ for real $M$ smaller than the 
saturation value, at leading order in the $1/N$ expansion.

\begin{figure}
\hspace{1.5cm}
    \epsfxsize = 15cm
    \epsffile{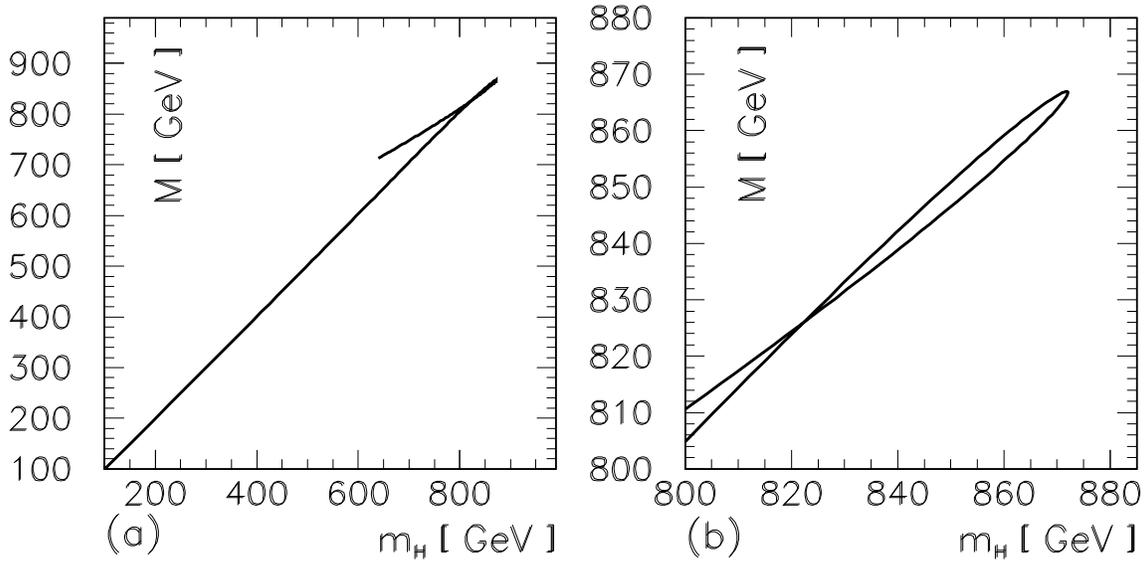}
\caption{{\em The nonperturbative relation between the Higgs pole mass $M$ and 
         the on--shell mass $m_H$ which results from the
         $1/N$ expansion at leading order. No singularity exists in
         this relation at this order, for real values of $M$ in the allowed
         range under the saturation point.}}
\end{figure}

The important difference between the leading order $1/N$ solution of the Higgs
sector and the model of ref. \cite{beenakker}, which exhibits the pathological
behaviour of the on--shell mass we are looking for, is that the self--energy
of the latter contains linear terms in $s$, which are absent in eq. 12.
This is related to the absence of a contribution to the Higgs wave function
renormalization at leading order in the $O(N)$ model. However,
contributions of this type are present at next--to--leading order in the $1/N$
expansion.

It is difficult to find out if the linear terms which should appear
at NLO in the $1/N$ expansion are large enough to induce the branching
point in the $m_H=m_H(M)$ relation we are looking for, without actually
performing a NLO calculation of the Higgs propagator. This is unfortunately 
a very challenging task. 

Still, one can make a guess about the size of
higher order contributions in the $1/N$ expansion by comparing the leading
order with the perturbative result.
As one can see in fig. 2, the perturbative results and the LO $1/N$ result
differ considerably already for a Higgs mass of the order of 600 GeV.
At these values of the Higgs mass perturbation theory appears to be
well under control. The convergence in both on--shell and pole schemes is good,
and the two schemes agree very well at NNLO. 
In fact, the NNLO perturbative results
in the pole and on--shell scheme agree well  
up to about 900 GeV. Unless one takes the view that there is
something fundamentally wrong with either the $1/N$ expansion, or perturbation
theory, or both, this discrepancy suggests that the ${\cal O}(1/N)$ 
corrections are numerically rather substantial.


\section{Conclusions}

Perturbation theory at the Higgs energy scale diverges very badly
at about 1 TeV in the on--shell scheme. In particular,
the two--loop correction to the Higgs width exceeds the one--loop correction
if the on--shell Higgs mass is larger than 930 GeV. For a Higgs boson
in this mass range the prediction of the position of the Higgs 
pole is rather unreliable.

We show that the pole renormalization scheme has much better convergence
properties. In this scheme, the two--loop corrections to the width become
as large as the one--loop ones only at 1.74 GeV. This suggests that the pole
renormalization scheme is preferable for describing a heavy Higgs boson
in phenomenological studies of heavy Higgs production at future colliders.

This choice of renormalization scheme cannot be justified in the same way
one chooses the renormalization scale in QCD for resuming large logarithms
in higher orders. However, this different behaviour with respect 
to the convergence
range of the on--shell scheme versus the pole scheme may be
the result of the existence of a relation between $m_H$ and $M$ at the
nonperturbative level. The observed convergence properties are consistent
with the assumption that the function $m_H=m_H(M)$ has a singularity at 
an energy of the order of 1 TeV, and beyond this critical value the on--shell 
mass is ill--defined.

We examine the leading order nonperturbative solution of the 
${\cal O}(N)$ model in the $1/N$ expansion. No branching point is present 
in the $m_H=m_H(M)$ function in this approximation for real values of $M$. 
The rather large 
deviation of the leading order $1/N$ expansion from the perturbative
result in the range where the latter is expected to be accurate leaves
room for substantial ${\cal O}(1/N)$ corrections. At next--to--leading
order in the $1/N$ expansion, the Higgs self--energy $\Sigma(s)$ 
is expected to acquire linear terms in $s$, similarly to the model studied
in ref. \cite{beenakker}, which exhibits a branching point.
Whether the ${\cal O}(1/N)$ corrections induce indeed a branching point in the
$m_H=m_H(M)$ relation at about 1 TeV, is an open question.


\pagebreak

\vspace{.5cm}

{\bf Acknowledgements}

We are indebted to Scott Willenbrock, Jochum van der Bij, and George Jikia
for useful discussions.
One of us (A. G.) is grateful 
to Stanley Brodsky for very interesting discussions, 
and would also like to thank 
the theory department of Brookhaven National Laboratory for its hospitality, 
and the US Department of Energy (DOE) for support. 
T. B. gratefully aknowledges the hospitality
of CERN, where part of the work was done. The work of A. G. was supported 
by the Deutsche Forschungsgemeinschaft (DFG).


\newpage


\end{document}